\documentclass[a4paper,fullpage]{article}

\usepackage[a4paper, textwidth=15.0cm]{geometry}
\usepackage{url}
\usepackage{xcolor}
\usepackage{hyperref}

\usepackage{graphicx}
\usepackage{subcaption}
\usepackage{amsmath, bm}
\usepackage{bbold}
\usepackage{cite}

\usepackage{authblk}
\usepackage{hyperref}

\usepackage{fancyhdr}
\newcommand*{\Bm}[1]{\boldsymbol{#1}}
\newcommand*{\Bv}[1]{\boldsymbol{#1}}

\newcommand{\CAPM}{single-index model}

\newcommand{\E}{\mathbb{E}}
\newcommand{\rowof}[2]{{#1}_{{#2},:}}
\newcommand{\colof}[2]{{#1}_{:,{#2}}}

\newcommand*{\citep}[1]{\cite{#1}}
\newcommand*{\citet}[1]{\cite{#1}}

\title{Applications of Gaussian Process Latent Variable Models in Finance
\footnote{This is a pre-print of an article accepted at the IntelliSys 2019.}}

\author[1,2]{Rajbir-Singh Nirwan\thanks{nirwan@fias.uni-frankfurt.de}}
\author[1,2]{Nils Bertschinger\thanks{bertschinger@fias.uni-frankfurt.de}}

\affil[1]{Frankfurt Institute for Advanced Studies, Frankfurt am Main, Germany}
\affil[2]{Goethe University, Frankfurt am Main, Germany}

\begin{document}
\maketitle       

\begin{abstract}
Estimating covariances between financial assets plays an important
role in risk management. In practice,
when the sample size is small compared to the number of variables, 
the empirical estimate is known to be very unstable.
Here, we propose a novel covariance estimator based on the Gaussian
Process Latent Variable Model (GP-LVM). Our estimator can be
considered as a non-linear extension of standard factor models with
readily interpretable parameters reminiscent of market
betas. Furthermore, our Bayesian treatment naturally shrinks the
sample covariance matrix towards a more structured matrix given by the
prior and thereby systematically reduces estimation errors.
Finally, we discuss some financial applications of the GP-LVM.
\end{abstract}

\section{Introduction}

Many financial problems require the estimation of covariance matrices
between given assets.  This may be useful to optimize one's portfolio,
i.e.: maximize the portfolio returns $\Bm{w}^T\Bm{r}$ and/or minimize
the volatility $\sqrt{\Bm{w}^T\Bm{K}\Bm{w}}$. Indeed, Markowitz
received a Noble Price in economics for his treatment of modern
portfolio theory \citep{Markowitz1952}.  In practice, estimating
historical returns and high-dimensional covariance matrices is
challenging and often times equally weighted portfolio outperforms the
portfolio constructed from sample estimates \citep{Jobson1981}.
The estimation of covariance matrices is especially hard, when the number 
of assets is large compared to the number of observations. Sample 
estimations in those cases are very unstable or can even become 
singular. To cope with this problem, a wide range of estimators, 
e.g. factor models such as the \CAPM ~\citep{SharpeCAPM} or shrinkage 
estimators \citep{Ledoit2004}, have been developed and 
employed in portfolio optimization. 

With todays machine learning techniques we can even further improve those estimates. 
Machine learning has already arrived in finance. Nevmyvaka et al. \citet{Nevmyvaka2006} 
trained an agent via 
reinforcement learning to optimally execute trades. 
Gately \citet{Gately1995} forecast asset prices with neural networks 
and Chapados et al. \citet{Chapados2008} with Gaussian Processes. Recently, 
Heaton et al. \citet{Heaton2016} made an ansatz to optimally allocate 
portfolios using deep autoencoders.
Wu et al. \citet{Wu2014} used Gaussian Processes to build volatility models and
Wilson et al. \citet{Wilson2011} to estimate time varying covariance matrices.
Bayesian machine learning methods are used more and more in this domain.
The fact, that in a Bayesian framework parameters are not treated as true
values, but as random variables, accounts for estimation uncertainties and can 
even alleviate the unwanted impacts of outliers. Furthermore, one can easily 
incorporate additional information and/or personal views by selecting suitable priors. 

In this paper, we propose a Bayesian covariance estimator based on the 
Gaussian Process Latent Variable Models (GP-LVMs) \citep{Lawrence2005},
which can be considered as a non-linear extension of standard 
factor models with readily interpretable parameters reminiscent of 
market betas. Our  Bayesian treatment naturally 
shrinks the sample covariance matrix (which maximizes the likelihood 
function) towards a more structured matrix given by the prior and 
thereby systematically reduces estimation errors.
We evaluated our model on the stocks of S\&P500 and found 
significant improvements in terms of model fit  
compared to classical linear models. 
Furthermore we suggest some financial applications, where Gaussian Processes can
be used as well. That includes portfolio allocation, price prediction for less 
frequently traded stocks and non-linear clustering of stocks into their sub-sectors.

In section \ref{background} we begin with an introduction to the Bayesian 
non-parametric Gaussian Processes and discuss the associated requirements for 
learning. Section \ref{finance} introduces the financial background
needed for portfolio optimization and how to relate it to Gaussian Processes.
In section \ref{experiments} we conduct experiments on covariance matrix 
estimations and discuss the results. We conclude in section \ref{conclusion}.

\section{Background}
\label{background}

In this paper, we utilize a Bayesian non-parametric machine learning 
approach based on Gaussian Processes (GPs). Combining those with 
latent variable models leads to Gaussian Process Latent Variable Models
(GP-LVMs), that we use to estimate the covariance between different
assets. These approaches have been described in detail in
\citep{Lawrence2005, Rasmussen2006}.
We provide a brief review here. Subsequently, we show, how to relate 
those machine learning approaches to the known models in finance, e.g.
the \CAPM ~\citep{SharpeCAPM}.

\subsection{Gaussian Processes}

A Gaussian Process (GP) is a generalization of the Gaussian distribution.
Using a GP, we can define a distribution over functions $f(\Bv{x})$,
where $\Bv{x} \in \mathbb{R}^Q$ and $f(\cdot) \in \mathbb{R}$.
Like a Gaussian distribution, the GP is specified by a mean
and a covariance. In the GP case, however, the mean is a function of the input
variable $m(\Bv{x})$ and the covariance is a function of two variables 
$k(\Bv{x}, \Bv{x}')$, which contains information about how the GP 
evaluated at $\Bv{x}$ and $\Bv{x}'$ covary
\begin{align}
 m(\Bv{x}) &= \mathbb{E}[f(\Bv{x})], \\
 k(\Bv{x}, \Bv{x}') &= \text{cov}(f(\Bv{x}), f(\Bv{x}')).
\end{align}
We write $f \sim \text{GP}(m(\cdot), k(\cdot, \cdot))$.
Any finite collection of function values, at $\Bv{x}_1, \ldots, \Bv{x}_N$, is 
jointly Gaussian distributed
\begin{equation}
 p(f(\Bv{x}_1), f(\Bv{x}_2), ... ,f(\Bv{x}_N)) = \mathcal{N}(\Bv{\mu}, \Bm{K}),
 \label{jgd}
\end{equation}
where $\Bv{\mu} = (m(\Bv{x}_1), ... ,m(\Bv{x}_N))^T$ is the mean vector and
$\Bm{K} \in \mathbb{R}^{N \times N}$ is the Gram matrix with entries $\Bm{K}_{ij} = 
k(\Bv{x}_i, \Bv{x}_j)$.
We refer to the covariance function as kernel function. The properties of the
function $f$ (i.e. smoothness, periodicity) are determined by the choice of this
kernel function. For example, sampled functions from a GP with an exponentiated 
quadratic covariance function $k_{\text{SE}}(\Bv{x}, \Bv{x}') = 
\sigma^2 \exp(-0.5 ||\Bv{x}- \Bv{x}'||^2_2/l^2)$ smoothly vary with lengthscale 
$l$ and are infinitely often differentiable.

Given a dataset $\mathcal{D}$ of $N$ input points $\Bm{X} = (\Bv{x_1}, ..., 
\Bv{x_N})^T$ and $N$ corresponding targets $\Bv{y} = ( y_1, ..., y_N)^T$,
the predictive distribution for a zero mean GP at $N_*$ new locations $\Bv{X}_*$
reads \citep{Rasmussen2006}
\begin{equation}
 \Bv{y}_* | \Bm{X}_*, \Bv{y}, \Bm{X} \sim \mathcal{N}(\Bv{f}_*, \Bm{K}_* ),
\end{equation}
where
\begin{align}
 \Bv{f}_* &= \Bm{K}_{\Bm{X}_*\Bm{X}}\Bm{K}_{\Bm{XX}}^{-1}\Bv{y}, \label{gpmean} \\
 \Bm{K}_* &= \Bm{K}_{\Bm{X}_*\Bm{X}_*} 
 - \Bm{K}_{\Bm{X}_*\Bm{X}}\Bm{K}_{\Bm{XX}}^{-1}\Bm{K}_{\Bm{X}\Bm{X}_*}. \label{gpcov}
\end{align}
$\Bm{K}_{\Bm{X}_*\Bm{X}} \in \mathbb{R}^{N_* \times N}$ is the covariance matrix
between the GP evaluated at $\Bm{X}_*$ and $\Bm{X}$, $\Bm{K}_{\Bm{XX}} \in 
\mathbb{R}^{N \times N}$ is the covariance matrix of the GP evaluated at $\Bm{X}$.
As we can see in equations (\ref{gpmean}) and (\ref{gpcov}), the
kernel function plays a very important role in the GP framework and will 
be important for our financial model as well.

\subsection{Gaussian Process Latent Variable Model}

Often times we are just given a data matrix $\Bm{Y} \in \mathbb{R}^{N \times D}$ 
and the goal is to find a lower dimensional representation $\Bm{X} 
\in \mathbb{R}^{N \times Q}$, without 
losing too much information. Principal component analysis (PCA) is one of the 
most used techniques for reducing the dimensions of the data, which has 
also been motivated as the maximum 
likelihood solution to a particular form of Gaussian Latent Variable 
Model \citep{Tipping1999}. PCA embeds $\Bm{Y}$ via a linear mapping 
into the latent space $\Bm{X}$. Lawrence \citet{Lawrence2005} introduced 
the Gaussian Process Latent Variable Model (GP-LVM) as a non-linear 
extension of probabilistic PCA. The generative procedure takes the form
\begin{equation}
 \rowof{\Bm{Y}}{n} = \Bv{f}(\rowof{\Bm{X}}{n}) + \Bv{\epsilon}_n,
 \label{gp1}
\end{equation}
where $\Bv{f} = (f_1, ... , f_D)^T$ is a group of $D$ independent samples from
a GP, i.e. $f_d \sim \text{GP}(0, k(\cdot, \cdot))$.
By this, we assume the rows of $\Bm{Y}$
to be jointly Gaussian distributed  and the
columns to be independent, i.e. each sample 
$\colof{\Bm{Y}}{d} \sim \mathcal{N}(\colof{\Bv{Y}}{d}| \Bv{0}, \Bm{K})$
where $\Bm{K} = k(\Bm{X}, \Bm{X}) +\sigma^2 \mathbf{I}$ and $\sigma^2$ 
denotes the variance of the random noise $\Bm{\epsilon}$.
The marginal likelihood of $\Bm{Y}$ becomes \citep{Lawrence2005}
\begin{equation}
 p(\Bm{Y}|\Bm{X}) = \prod_{d=1}^D \mathcal{N}(\colof{\Bv{Y}}{d}| \Bv{0}, 
 \Bm{K}) = \frac{1}{(2\pi)^{ND/2}|\Bm{K}|^{D/2}} \exp \left( -\frac{1}{2}
 \text{tr}(\Bm{K}^{-1}\Bm{YY}^T) \right).
 \label{marginalLikelihood}
\end{equation}
The
dependency on the latent positions $\Bm{X}$ and the kernel hyperparameters
is given through the kernel matrix $\Bm{K}$. As suggested by 
Lawrence \citet{Lawrence2005}, we can optimize the log marginal likelihood 
$\log p(\Bm{Y}|\Bm{X})$ with respect to the latent positions
and the hyperparameters.

\subsection{Variational Inference}
\label{vb}

Optimization can easily lead to overfitting. Therefore, a fully Bayesian
treatment of the model would be preferable but is intractable. 
Bishop \citet{Bishop2006} explains the variational 
inference framework, which not only handles the problem of overfitting
but also allows to automatically select the dimensionality of the
latent space. 
Instead of optimizing equation (\ref{marginalLikelihood}), 
we want to calculate the posterior using Bayes rule $p(\Bm{X}|\Bm{Y}) = 
p(\Bm{Y}|\Bm{X})p(\Bm{X})/p(\Bm{Y})$, which is intractable.
The idea behind variational Bayes is to approximate the true posterior
$p(\Bm{X}|\Bm{Y})$ by another distribution $q(\Bm{X})$, selected from a
tractable family. The goal is to select the one distribution
$q(\Bm{X})$, that is closest to the true posterior $p(\Bm{X}|\Bm{Y})$ in
some sense. A natural choice to quantify the closeness is given by the
Kullback-Leibler divergence \citep{Cover1991}
\begin{equation}
 \text{KL}[q(\Bm{X})||p(\Bm{X}|\Bm{Y})] = \int q(\Bm{X}) 
 \log \frac{q(\Bm{X})}{p(\Bm{X}|\Bm{Y})} \mathrm{d}\Bm{X}. 
 \label{klDivergence}
\end{equation}

By defining $\tilde{p}(\Bm{X}|\Bm{Y}) = p(\Bm{Y}|\Bm{X})p(\Bm{X})$ as the
unnormalized posterior, equation (\ref{klDivergence}) becomes
\begin{align}
 \text{KL}[q(\Bm{X})||p(\Bm{X}|\Bm{Y})] &= \int q(\Bm{X}) 
 \log \frac{q(\Bm{X})}{\tilde{p}(\Bm{X}|\Bm{Y})} \mathrm{d}\Bm{X} +
 \log p(\Bm{Y}) \nonumber \\
 &= - \underbrace{\mathbb{E}_{q(\Bm{X})} \left[ \log 
 \frac{\tilde{p}(\Bm{X}|\Bm{Y})}{q(\Bm{X})} \right]}_{\text{ELBO}} +
 \log p(\Bm{Y})
 \label{elboEq}
\end{align}
with the first term on the right hand side being known as the evidence lower
bound (ELBO).
Equation (\ref{elboEq}) is the objective function we want to minimize
with respect to $q(\Bm{X})$ to get a good approximation to the true posterior.
Note that on the left hand side only the ELBO is $q$ dependent. So, in order to
minimize (\ref{elboEq}), we can just as well maximize the ELBO. 
Because the Kullback-Leibler divergence is non-negative,
the ELBO is a lower bound on the evidence $\log p(\Bm{Y})$\footnote{The evidence 
$\log p(\Bm{Y})$ is also referred to as log marginal
likelihood in the literature. The term marginal likelihood is already used for 
$p(\Bm{Y}|\Bm{X})$ in this paper. Therefore, we will refer to 
$\log p(\Bm{Y})$ as the evidence.}.
Therefore, this procedure not only gives the best approximation to the 
posterior within the variational family under the KL ceriterion, 
but it also bounds the evidence, 
which serves as a measure 
of the goodness of our fit. The number of latent dimensions $Q$ 
can be chosen to be the one, which maximizes the ELBO.

So, GP-LVM is a model, which reduces the dimensions of our data-matrix
$\Bm{Y} \in \mathbb{R}^{N \times D}$ from $D$ to $Q$ in a non-linear way and
at the same time estimates the covariance matrix between the $N$ points.
The estimated covariance matrix can then be used for further analysis.

\section{Finance}
\label{finance}
Now we have a procedure to estimate the covariance matrix between different datapoints. 
This section discusses how we can relate this to financial models.
\subsection{CAPM}
The Capital Asset Pricing Model (CAPM) describes the relationship between the expected returns
of an asset $\Bv{r}_n \in \mathbb{R}^{D}$ for D days and its risk $\beta_n$
\begin{equation}
 \E[\Bv{r}_n] = \Bv{r}_f + \beta_n \E[\Bv{r}_m - \Bv{r}_f],
 \label{capm}
\end{equation}
where $\Bv{r}_f \in \mathbb{R}^D$ is the risk free return on $D$ different days
and $\Bv{r}_m$ is the market return on D different days.
The main idea behind CAPM is, that an investor needs to be compensated for the risk of his 
holdings. For a risk free asset with $\beta_n=0$, the expected return $\mathbb{E}[\Bm{r}_n]$
is just the risk free rate $\Bv{r}_f$. 
If an asset is risky with a risk $\beta_n \neq 0$, the expected return
$\E[\Bv{r}_n]$ is increased by $\beta_n \E[\tilde{\Bv{r}}_m]$, where $\tilde{\Bv{r}}_m$ 
is the excess return of the market $\tilde{\Bv{r}}_m = \Bv{r}_m - \Bv{r}_f$. 

We can write down equation (\ref{capm}) in terms of the excess return 
$\tilde{\Bv{r}} = \Bv{r}-\Bv{r}_f$ and get
\begin{equation}
 \E[\tilde{\Bv{r}}_n] = \beta_n \E[\tilde{\Bv{r}}_m],
 \label{capm2}
\end{equation}
where $\tilde{\Bv{r}}_n$ is the excess return of a given asset and $\tilde{\Bv{r}}_m$ 
is the excess return of the market (also called a risk factor). 
Arbitrage pricing theory \citep{Ross1995} generalizes the above model by allowing multiple 
risk factors $\Bv{F}$ beside the market $\tilde{\Bv{r}}_m$. 
In particular, it assumes that asset returns follow a factor structure
\begin{equation}
 \Bv{r}_n = \alpha_n + \Bv{F} \Bv{\beta}_n + \Bv{\epsilon}_n,
 \label{apt}
\end{equation}
with $\Bv{\epsilon}_n$ denoting some independent zero mean noise with variance $\sigma^2_n$.
Here, $\Bm{F} \in \mathbb{R}^{D \times Q}$ is the matrix of Q factor returns on D days and 
$\Bv{\beta}_n \in \mathbb{R}^Q$ is the loading of stock $n$ to the Q factors.
Arbitrage pricing theory \citep{Ross1976} then shows that the expected excess returns adhere to
\begin{equation}
  \label{eq:apt_CAPM}
  \E[\tilde{\Bv{r}}_n] = \E[\tilde{\Bv{F}}] \Bv{\beta}_n,
\end{equation}
i.e. the CAPM is derived as special case when assuming a single risk factor (\CAPM).

To match the form of the GP-LVM (see equation (\ref{gp1})), we rewrite equation (\ref{apt}) as
\begin{equation}
 \rowof{\Bm{r}}{n} = \Bv{f}(\rowof{\Bm{B}}{n}) + \Bv{\epsilon}_n, 
 \label{gpf}
\end{equation}
where $\Bm{r} \in \mathbb{R}^{N \times D}$ is the return matrix\footnote{To stay 
consistent with the financial literature, we denote
  the return matrix with lower case $\Bm{r}$.}.  Note that assuming
latent factors distributed as $\Bv{F} \sim \mathcal{N}(0, 1)$ and
marginalizing over them, equation (\ref{apt}) is a special case of equation
(\ref{gpf}) with $\Bv{f}$ drawn from a GP mapping $\Bv{\beta}_n$ to $\Bm{r}_n$
with a linear kernel.
Interestingly, this provides an exact correspondence with factor
models by considering the matrix of couplings
$\Bm{B} = (\Bv{\beta_1}, ..., \Bv{\beta_N})^T$ as the latent space
positions\footnote{Because of the context, from now on we will use
  $\Bv{\beta}$ for the latent space instead of $\Bv{x}$.}.  In this perspective
the factor model can be seen as a linear dimensionality reduction,
where we reduce the $N \times D$ matrix $\Bm{r}$ to a low rank matrix
$\Bm{B}$ of size $N \times Q$.  By chosing a non-linear kernel
$k(\cdot , \cdot)$ the GP-LVM formulation readily allows for
non-linear dimensionality reductions.  Since, it is generally known,
that different assets have different volatilities, we further
generalize the model.  In particular, we assume the noise
$\Bm{\epsilon}$ to be a zero mean Gaussian, but allow for different
variances $\sigma^2_n$ for different stocks.  For this reason, we also
have to parameterize the kernel (covariance) matrix in a different way
than usual. Section \ref{dcam} explains how to deal with that.
The model is then approximated using variational inference as described in section \ref{vb}.
After inferring $\Bm{B}$ and the hyperparameters of the kernel, 
we can calculate the covariance matrix $\Bm{K}$ and use it for further analysis.

\subsection{Modern Portfolio Theory}
\label{modernPortfolioTheory}
Markowitz \citet{Markowitz1952} provided the foundation for
modern portfolio theory, for which he received a Nobel Prize in economics.
The method analyses how good a given portfolio is, based on the mean and 
the variance of the returns of the assets contained in the portfolio. This
can also be formulated as an optimization problem for selecting an optimal
portfolio, given the covariance between the assets and the risk tolerance 
$q$ of the investor. 

Given the covariance matrix $\Bm{K} \in \mathbb{R}^{N \times N}$, we can calculate the
optimal portfolio weights $\Bv{w}$ by
\begin{equation}
 \Bv{w}_{\text{opt}} = \min_{\Bv{w}} (\Bv{w}^T \Bm{K} \Bv{w} - q \bar{\Bv{r}}^T \Bv{w}),
 \label{markowitzPortfolio}
\end{equation}
where $\bar{\Bv{r}}$ is the mean return vector.
Risk friendly investors have a higher $q$ than risk averse investors.
The model is constrained by $\sum_{\Bv{w}} = 1$.
Since $\bar{\Bv{r}}$ is very hard to estimate in general and we are primarily
interested in the estimation of the covariance matrix $\Bm{K}$,
we set $q$ to zero and get
\begin{equation}
 \Bv{w}_{\text{opt}} = \min_{\Bv{w}}( \Bv{w}^T \Bm{K} \Bv{w} ).
 \label{minimalRiskPortfolio}
\end{equation}
This portfolio is called the minimal risk portfolio, i.e. the solution to equation
(\ref{minimalRiskPortfolio}) provides the weights for the portfolio, which minimizes
the overall risk, assuming the estimated $\Bm{K}$ is the true covariance matrix.

\section{Experiments}
\label{experiments}

In this section, we discuss the performance of the GP-LVM on financial data.
After describing the data collection and modeling procedure, 
we evaluate the model on the daily return series of the S\&P500 stocks. 
Subsequently, we discuss further financial applications. In particular, 
we show how to build a minimal risk portfolio
(this can easily be extended to maximizing returns as well), how to 
fill-in prices for assets which are not traded frequently and how
to visualize sector relations between different stocks (latent space embedding).

\subsection{Data Collection and Modeling}
\label{dcam}
For a given time period, we take all the stocks from the S\&P500, whose daily 
close prices were
available for the whole period\footnote{We are aware that this introduces survivorship bias. 
Thus, in some applications our results might be overly optimistic. Nevertheless, 
we expect relative comparisons between different models to be meaningful.}. 
The data were downloaded from Yahoo Finance.
After having the close prices in a matrix $\Bm{p}\in \mathbb{R}^{N \times (D+1)}$,
we calculate the return matrix $\Bm{r} \in \mathbb{R}^{N \times D}$,
where $\Bm{r}_{nd} =
(\Bm{p}_{n, d} - \Bm{p}_{n,d-1})/\Bm{p}_{n,d-1}$. $\Bm{r}$ builds
the basis of our analysis. 

We can feed $\Bm{r}$ into the GP-LVM. The GP-LVM procedure, as
described in section \ref{background}, assumes the
likelihood to be Gaussian with the covariance given by the kernel function
for each day and assumes independency over different days. We use 
the following kernel functions
\begin{align}
 &k_{\text{noise}}(\Bv{\beta}_i, \Bv{\beta}_j) = \sigma_{\text{noise},i}^2 \delta_{i,j}, \nonumber \\
 &k_{\text{linear}}(\Bv{\beta}_i, \Bv{\beta}_j) = \sigma^2 \Bv{\beta}_i^T 
 \Bv{\beta}_j,
\end{align}
and the stationary kernels
\begin{align}
 &k_{\text{se}}(\Bv{\beta}_i, \Bv{\beta}_j) =  k_{\text{se}}(d_{ij}) =  \exp 
 \left( -\frac{1}{2l^2} d_{ij}^2  \right), \nonumber \\
 &k_{\text{exp}}(\Bv{\beta}_i, \Bv{\beta}_j) = k_{\text{exp}}(d_{ij}) = \exp 
 \left( -\frac{1}{2l} d_{ij}  \right), \nonumber \\
 &k_{\text{m32}}(\Bv{\beta}_i, \Bv{\beta}_j) = k_{\text{m32}}(d_{ij}) =
 \left( 1 + \frac{\sqrt3 d_{ij} }{l} \right)
 \exp \left( -\frac{\sqrt3}{2l} d_{ij}  \right),
\end{align}
where $d_{ij} = ||\Bv{\beta}_i - \Bv{\beta}_j ||_2$ is the Euclidean distance 
between $\Bv{\beta}_i$ and $\Bv{\beta}_j$. $\sigma^2$ is the kernel variance
and $l$ kernel lengthscale. Note that since the diagonal elements of stationary kernels
are the same, they are not well suited for an estimation of a covariance matrix
between different financial assets. Therefore, in the case of stationary kernel we decompose our 
covariance matrix $\Bm{K}_{\text{cov}}$ into a vector of coefficient
scales $\Bv{\sigma}$ and a correlation matrix $\Bm{K}_{\text{corr}}$, such that
$\Bm{K}_{\text{cov}} = \Bm{\Sigma} \Bm{K}_{\text{corr}} \Bm{\Sigma}$, where 
$\Bm{\Sigma}$ is a diagonal matrix with $\Bv{\sigma}$ on the diagonal.
The full kernel function $k(\cdot, \cdot)$ at
the end is the sum of the noise kernel $k_{\text{noise}}$ and one of the 
other kernels. In matrix form we get
\begin{align}
 &\Bm{K}_{\text{linear}} = k_{\text{linear}}(\Bm{B},\Bm{B})+k_{\text{noise}}(\Bm{B},\Bm{B}), \nonumber \\ 
 &\Bm{K}_{\text{se}} = \Bm{\Sigma} \ k_{\text{se}}(\Bm{B},\Bm{B}) \ \Bm{\Sigma} + 
 k_{\text{noise}}(\Bm{B},\Bm{B}), \nonumber \\ 
 &\Bm{K}_{\text{exp}}=\Bm{\Sigma} \ k_{\text{exp}}(\Bm{B},\Bm{B}) \ \Bm{\Sigma} +
 k_{\text{noise}}(\Bm{B},\Bm{B}), \nonumber \\ 
 & \Bm{K}_{\text{m32}} = \Bm{\Sigma} \ k_{\text{m32}}(\Bm{B},\Bm{B}) \ \Bm{\Sigma} + 
 k_{\text{noise}}(\Bm{B},\Bm{B}),
 \label{covariance_construction}
\end{align}
where $\Bm{B} = (\Bv{\beta}_1, ..., \Bv{\beta}_N)^T$. We chose the following priors
\begin{align}
 \Bm{B} \sim \mathcal{N}(0,1) \qquad \quad
 l, \sigma \sim \text{InvGamma}(3, 1)   \qquad \quad
 \Bv{\sigma}, \sigma_{noise} \sim \mathcal{N}(0,0.5). 
\end{align}

The prior on $\Bm{B}$ determines how much space is allocated to 
the points in the latent space. The volume of this space expands with higher
latent space dimension, which make the model prone to overfitting. 
To cope with that, we assign an inverse gamma
prior to the lengthscale $l$ and $\sigma$ ($\sigma$ in the 
linear kernel has a similar functionality as $l$ in the stationary kernels).
It allows larger values for higher latent space dimension, thereby shrinking 
the effective latent space volume and 
exponentially suppresses very small values, which deters overfitting as well.
The parameters of the priors are chosen such that they allow for enough volume
in the latent space for roughly 100-150 datapoints, which we use in our analysis.
If the number of points is drastically different, one should adjust the
parameters accordingly. The kernel standard deviations $\sigma_{\text{noise}}$ and
$\Bv{\sigma}$ are assigned a half Gaussian prior with variance 0.5, which
is essentially a flat prior, since the returns are rarely above 0.1 for a day.

Model inference under these specifications (GP-LVM likelihood for the data
and prior for $\Bm{B}$ and all kernel hyperparameters $\sigma, \sigma_{\text{noise}}, l$
and $\Bv{\sigma}$, which we denote by
$\Bm{\theta}$) was carried out by variational inference as described in section \ref{vb}.
To this end, we implemented all models
in the probabilistic programming language
Stan \citep{stan}, which supports variational inference out of the box.
We approximate the posterior by independent Gaussians.
The source code is available on Github\footnote{https://github.com/RSNirwan/GPLVMsInFinance}.
We tested different initializations for the parameter (random, PCA solution and 
Isomap solution), but there were no significant differences in the ELBO. 
So, we started the inference 50 times with random initializations and took the
result with highest ELBO for each kernel function and $Q$.

\subsection{Model Comparison}
\begin{figure}
 \centering
  \begin{subfigure}[h]{0.47\textwidth}
   \includegraphics[width=0.99\textwidth]{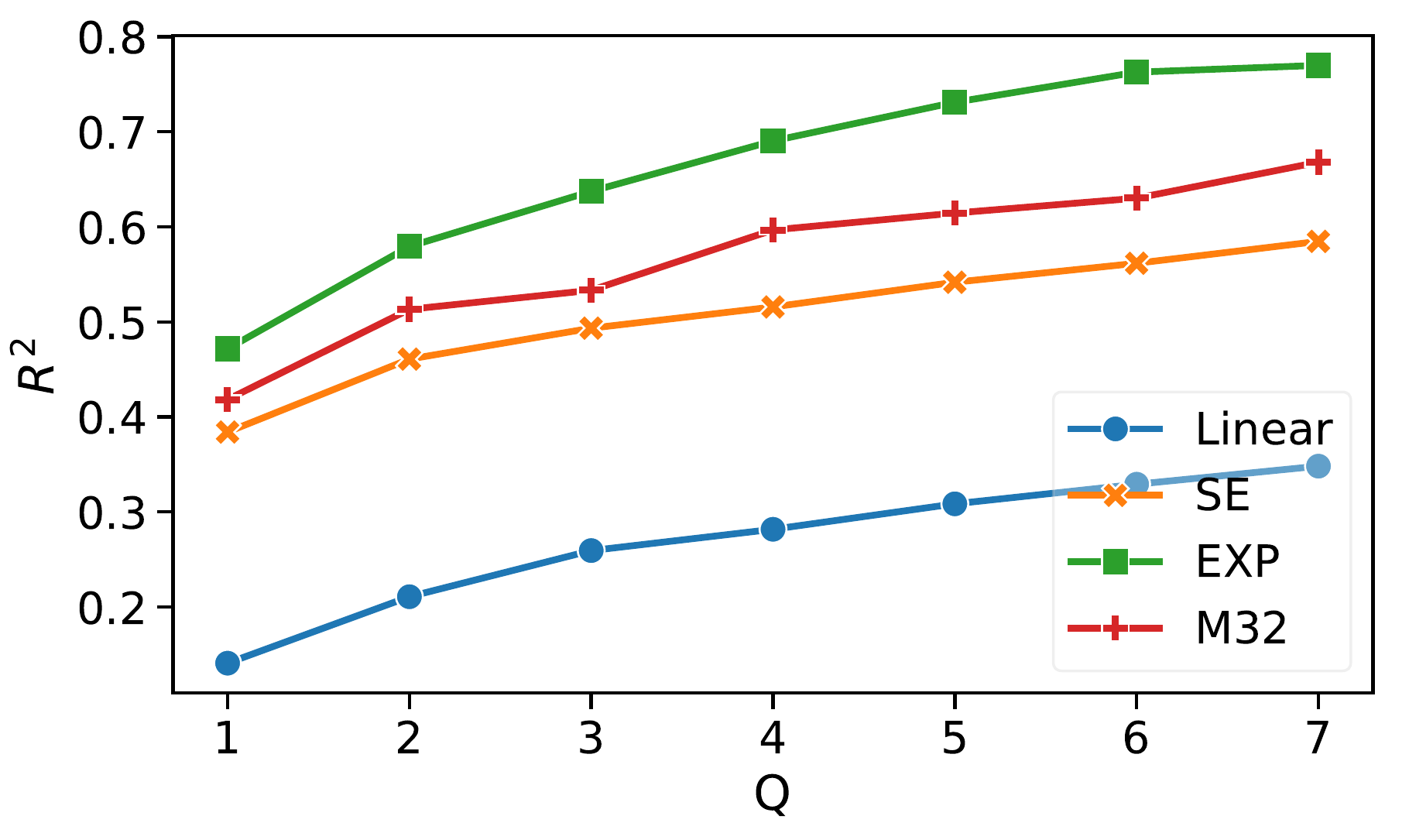}
  \end{subfigure}
  \quad
  \begin{subfigure}[h]{0.47\textwidth}
   \includegraphics[width=0.99\textwidth]{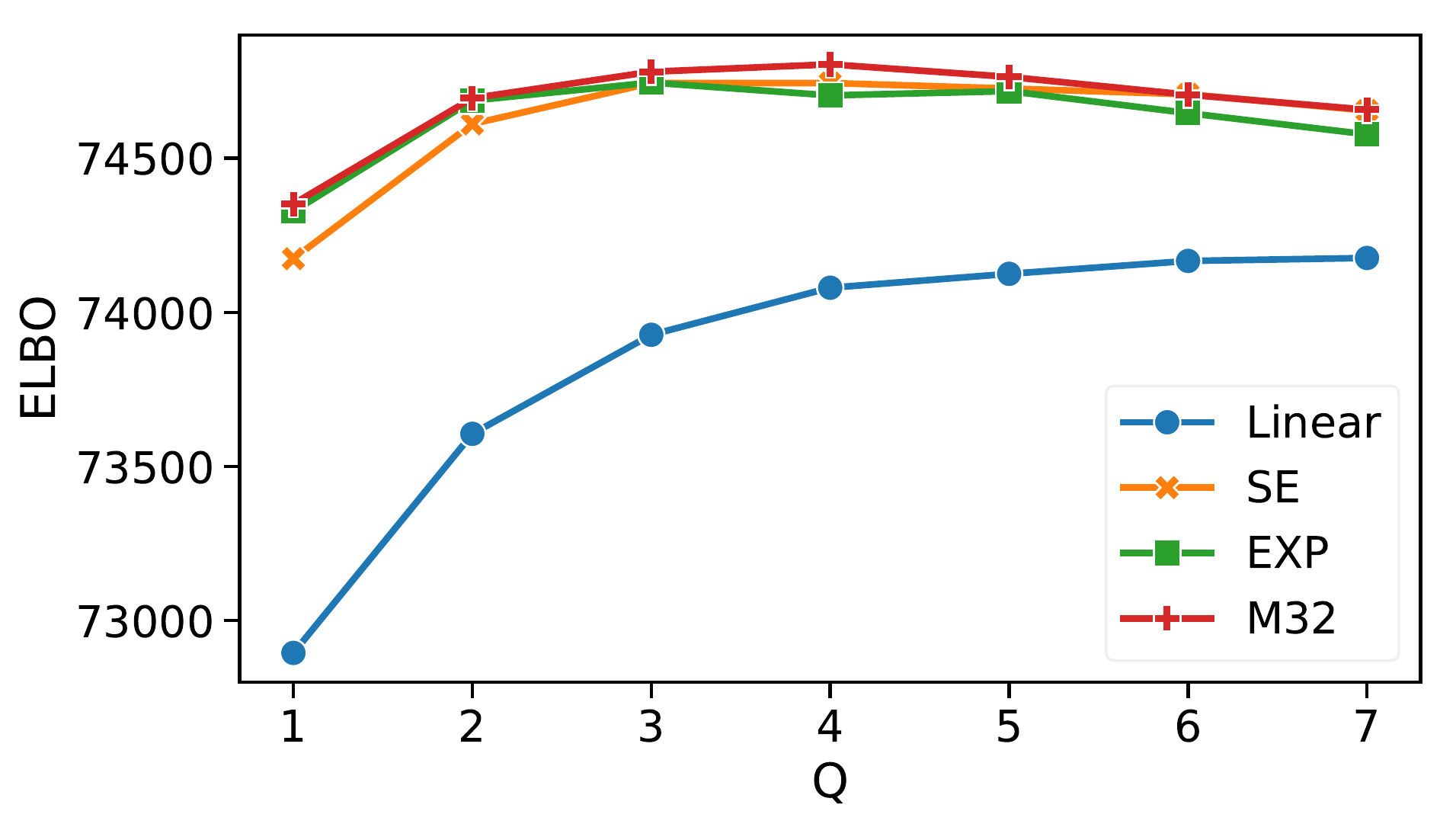}
  \end{subfigure}
 \caption{Left: $R^2$-score as a function of the latent dimension $Q$ for different 
 kernel functions. Right: ELBO as a function of the latent dimension $Q$. We randomly 
 chose 120 stocks from the S\&P500 and made the analysis on returns from Jan 2017 to 
 Dec 2017.}
 \label{r2_elbo}
\end{figure}

The GP-LVM can be evaluated in many different ways.
Since, it projects the data from a $D$-dimensional space to a $Q$-dimensional 
latent space, we can look at the reconstruction error. A suitable measure of
the reconstruction error is the R-squared ($R^2$) score, which is equal to one if there
is no reconstruction error and decreases if the reconstruction error increases.
It is defined by
\begin{equation}
 R^2 = 1- \frac{\sum_i(y_i - f_i)^2}{\sum_i (y_i - \bar{y})^2},
 \label{r2score}
\end{equation}
where $y = (y_1,...,y_N)^T$ are the true values, $f = (f_1, ... f_N)^T$
are the predicted values and $\bar{y}$ is the mean of $y$.
In the following, we look at the $R^2$ as a function of the latent dimension
$Q$ for different kernels. Figure \ref{r2_elbo} (left plot) shows the results for three 
non-linear kernels and the linear kernel. Only a single dimension in the 
non-linear case can already 
capture almost 50\% of the structure, whereas the linear kernel is at 15\%.
As one would expect, the higher $Q$, the more structure can be learned. 
But at some point the model will also start learning the noise and overfit.

As introduced in section \ref{vb}, the ELBO is 
a good measure to evaluate different models and already incorporates
models complexity (overfitting). Figure \ref{r2_elbo} (right plot) shows the ELBO
as a function of the latent dimension $Q$. Here we see, that the model selects
just a few latent dimensions. Depending on the used kernel, latent dimensions
from three to five are already enough to capture the structure. If we increase the
dimensions further, the ELBO starts dropping, which is a sign of overfitting.
As can be seen from Figure \ref{r2_elbo}, we do not need to go to higher dimensions. $Q$ 
between two and five is already
good enough and captures the structure that can be captured by the model.

\subsection{Applications}

The GP-LVM provides us the covariance matrix $\Bm{K}$ and the
latent space representation $\Bm{B}$ of the data. We can build a lot of nice 
applications upon that, some of which are discussed in this section.

\subsubsection{Portfolio Allocation}

After inferring $\Bm{B}$ and $\Bm{\theta}$,
we can reconstruct the covariance matrix $\Bm{K}$ using  
equation (\ref{covariance_construction}). Thereafter, we only need to minimize 
equation (\ref{minimalRiskPortfolio}), which provides the weights $\Bv{w}$ 
for our portfolio in the future. Minimization of (\ref{minimalRiskPortfolio})
is done under the constraints: $\sum_n \Bv{w}_n = 1$ and 
$0 < \Bm{w}_n < 0.1, \forall n$. These constraints are commonly employed in practice and
ensure that we are fully invested, take on long positions only and 
prohibit too much weight for a single asset.

For our tests, we proceed as follows: First, we get data
for 60 randomly selected stocks from the 
S\&P500 from Jan 2008 to Jan 2018. Then, we learn $\Bv{w}$ from the past year and
buy accordingly for the next six months. 
Starting from Jan 2008, this procedure is repeated every six months.
We calculate the average return, standard deviation and the Sharpe ratio.
\citet{Sharpe1966} suggested the Sharpe ratio as a measure for a portfolio`s performance,
which is the average return earned per unit of volatility and can be calculated 
by dividing the mean return of a series by its standard deviation.
Table \ref{table_sharpe}
shows the performance of the portfolio for different kernels for $Q = 3$.
For the GP-LVM we chose the linear, SE, 
EXP and M32 kernels and included the performance given by the sample 
covariance matrix, i.e. $\Bm{K} =\frac{1}{D} (\Bv{r} - \hat{\Bv{\mu}}) 
(\Bv{r} - \hat{\Bv{\mu}})^T$, where $\hat{\Bv{\mu}}_n = \frac{1}{D} \sum_{d = 1}^D \Bv{r}_{nd}$,
the shrunk Ledoit-Wolf covariance matrix\footnote{Here, we have used the implementation in the
Python toolbox scikit-learn \citep{scikit-learn}.} \citep{Ledoit2004} and the equally 
weighted portfolio, where $\Bv{w} = (1,1,...,1)/N$.

\begin{table}[ht!]
  \caption{Mean returns, standard deviation and the Sharpe ratio of 
    different models on a yearly basis.}
    \vspace{-0.2cm}
  \label{table_sharpe}
  \begin{center}
  \begin{tabular}{ l c c c c c c c }
      \hline 
      {Model} & {Linear} & {SE} & {EXP} & {M32} & 
      {Sample Cov} & {Ledoit Wolf} & {Eq. Weighted} \\[2pt] \hline
      Mean & 0.142 & 0.151 & 0.155 & 0.158 & 0.149 & 0.148 & 0.182 \\ 
      Std & 0.158 & 0.156 & 0.154 & 0.153 & 0.159 & 0.159 & 0.232 \\ 
      Sharpe ratio & 0.901 & 0.969 & 1.008 & 1.029 & 0.934 & 0.931 & 0.786 \\[2pt] \hline
    \end{tabular}
  \end{center}
\end{table}

Non-linear kernels have the minimal variance
and at the same time the best Sharpe ratio values. 
Note that we are building a minimal variance portfolio and therefore not maximizing
the mean returns as explained in section \ref{modernPortfolioTheory}. 
For finite $q$ in equation (\ref{markowitzPortfolio}) one can also build
portfolios, which not only minimize risk but also maximize the returns.
Another requirement for that would be to have a good estimator for the 
expected return as well.

\subsubsection{Fill in Missing Values}

Regulation requires fair value assessment of all assets \citep{FASB2006}, including 
illiquid and infrequently traded ones.
Here, we demonstrate how the GP-LVM could be used to fill-in missing prices, e.g.
if no trading took place. For illustration purposes, 
we continue working with our daily close 
prices of stocks from the S\&P500, but the procedure can be applied
to any other asset class and time resolution.

\begin{figure}
 \centering
  \begin{subfigure}[h]{0.45\textwidth}
   \includegraphics[width=0.99\textwidth]{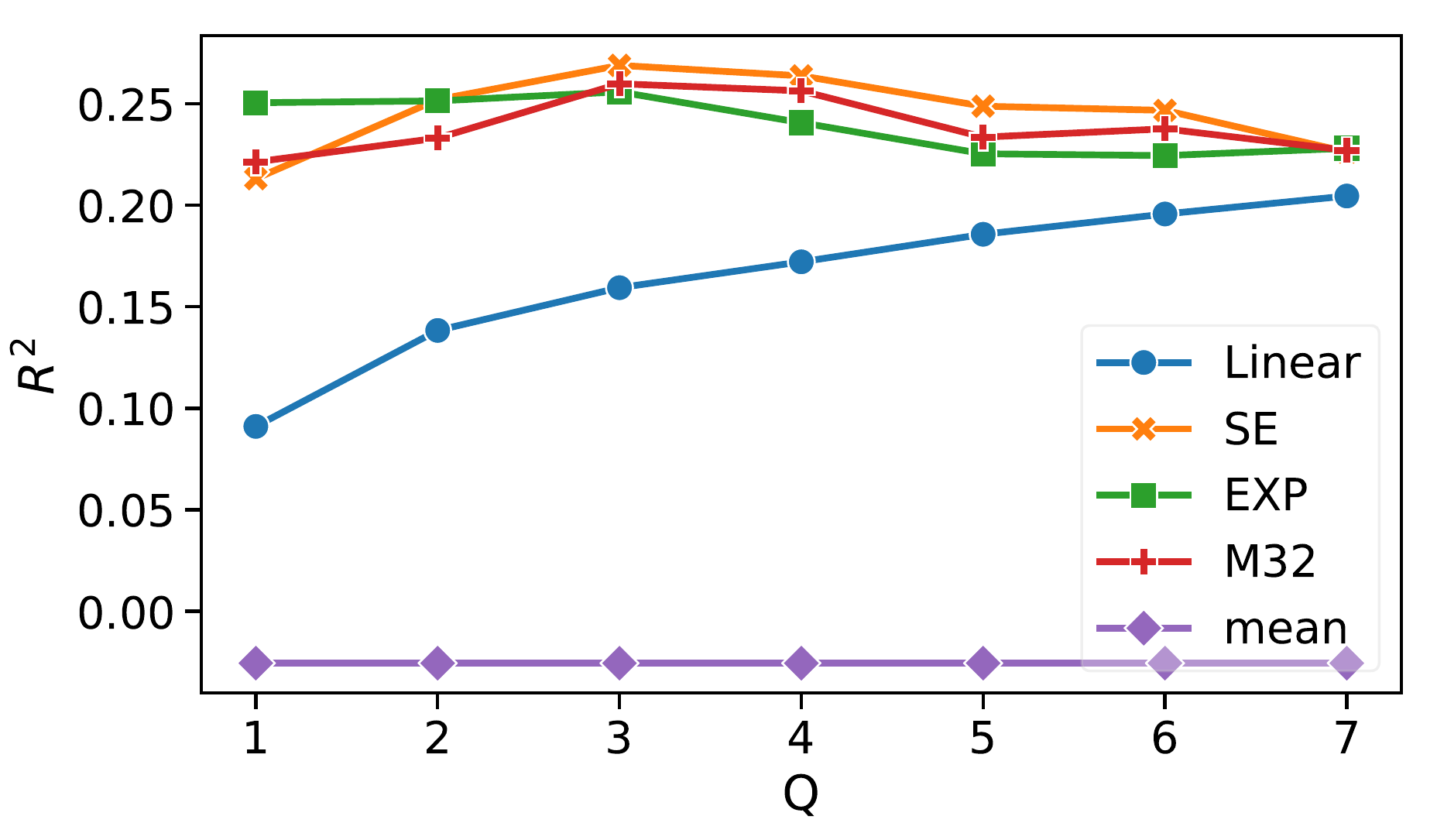}
  \end{subfigure}
  \quad
  \begin{subfigure}[h]{0.45\textwidth}
   \includegraphics[width=0.99\textwidth]{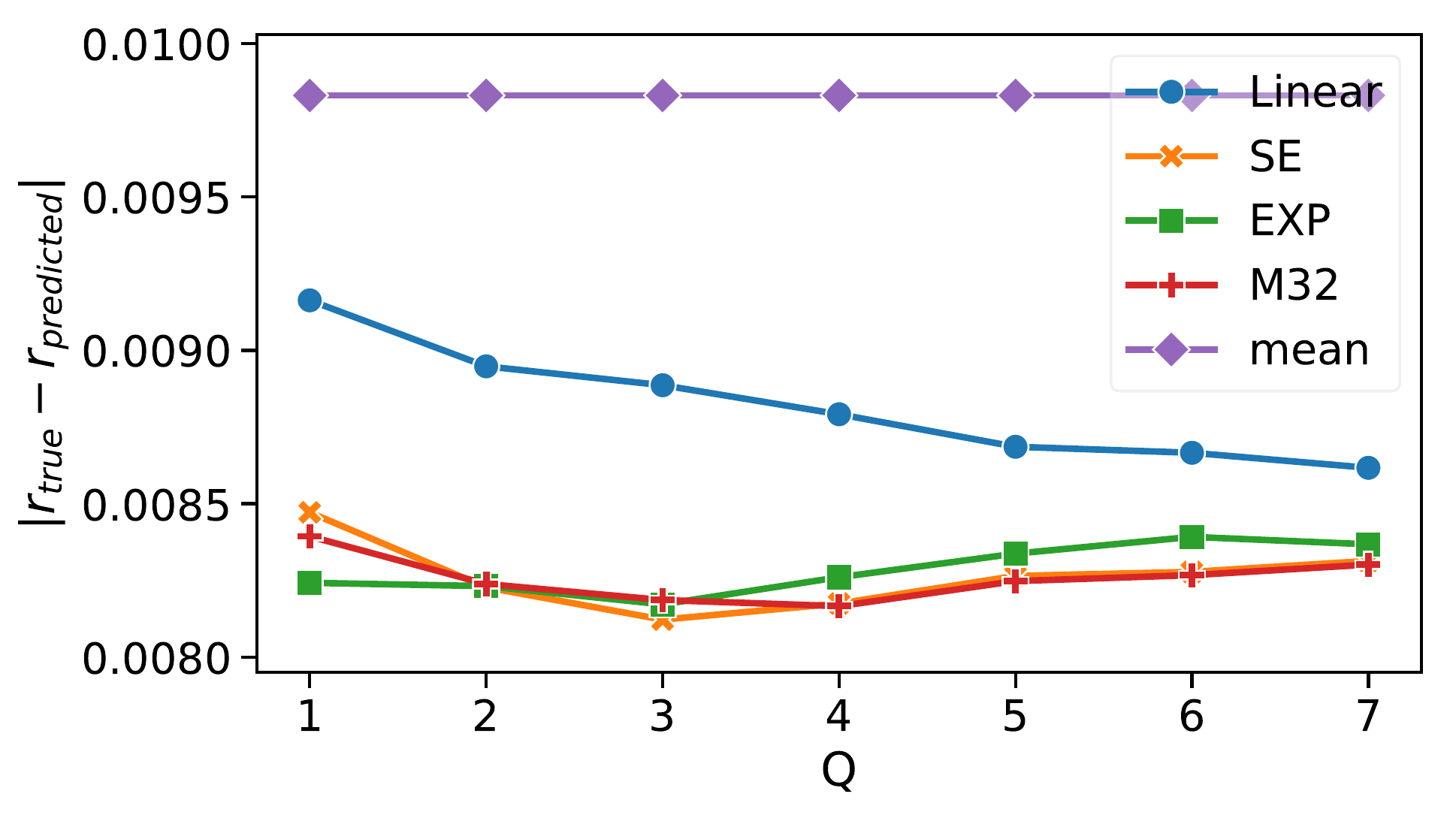}
  \end{subfigure}
 \caption{Left: $R^2$-score of the predicted values as a function of 
 the latent dimension $Q$. Right: The average absolute deviation of suggested return to the 
  real return evaluated by Leaving-one-out cross-validation.
  Historical mean is indicated by ``mean'' and is Q-independent.}
 \label{missing_values}
\end{figure}

First, we split our data $\Bm{r}$ into training and test set.
The test set contains days where the returns of assets are missing and our
goal is to accurately predict the return.
In our case, we use stock data from Jan 2017 to Oct 2017 for 
training and Nov 2017 to Dez 2017 to test.
The latent space $\Bm{B}$ and the hyperparameters $\Bm{\theta}$ are
learned from the training set. Given $\Bm{B}$ and $\Bv{\theta}$,
we can use the standard GP equations (eq. (\ref{gpmean}) and (\ref{gpcov})) 
to get the posterior return distribution. A suggestion for the value of the
return at a particular day can be made by the mean of this distribution.
Given $N$ stocks for a particular day $d$, we fit a GP to $N-1$ stocks
and predict the value of the remaining stock. We repeat the process
$N$ times, each time leaving out a different stock 
(Leave-one-out cross-validation). Figure \ref{missing_values} 
shows the $R^2$-score (eq. (\ref{r2score})) and the average absolute deviation 
$\frac{1}{ND}\sum_{nd}|\Bm{r}_{nd} - \Bm{r}^{\text{pred}}_{nd}|$
of the suggested return to the real return. The average is build 
over all stocks and all days in the test set.

The predictions with just the historical mean have a negative $R^2$-score.
The linear model is better than that. But if we switch to non-linear kernels, we
can even further increase the prediction score.
For $Q$ between 2 and 4 we obtain the best results.
Note that to make a decent suggestion for the return of an asset,
there must be some correlated assets in the data as well. Otherwise,
the model has no information at all about the asset, we want to predict 
the returns for.

\subsubsection{Interpretation of the Latent Space}

\begin{figure}
 \centering
  \includegraphics[width=0.65\textwidth]{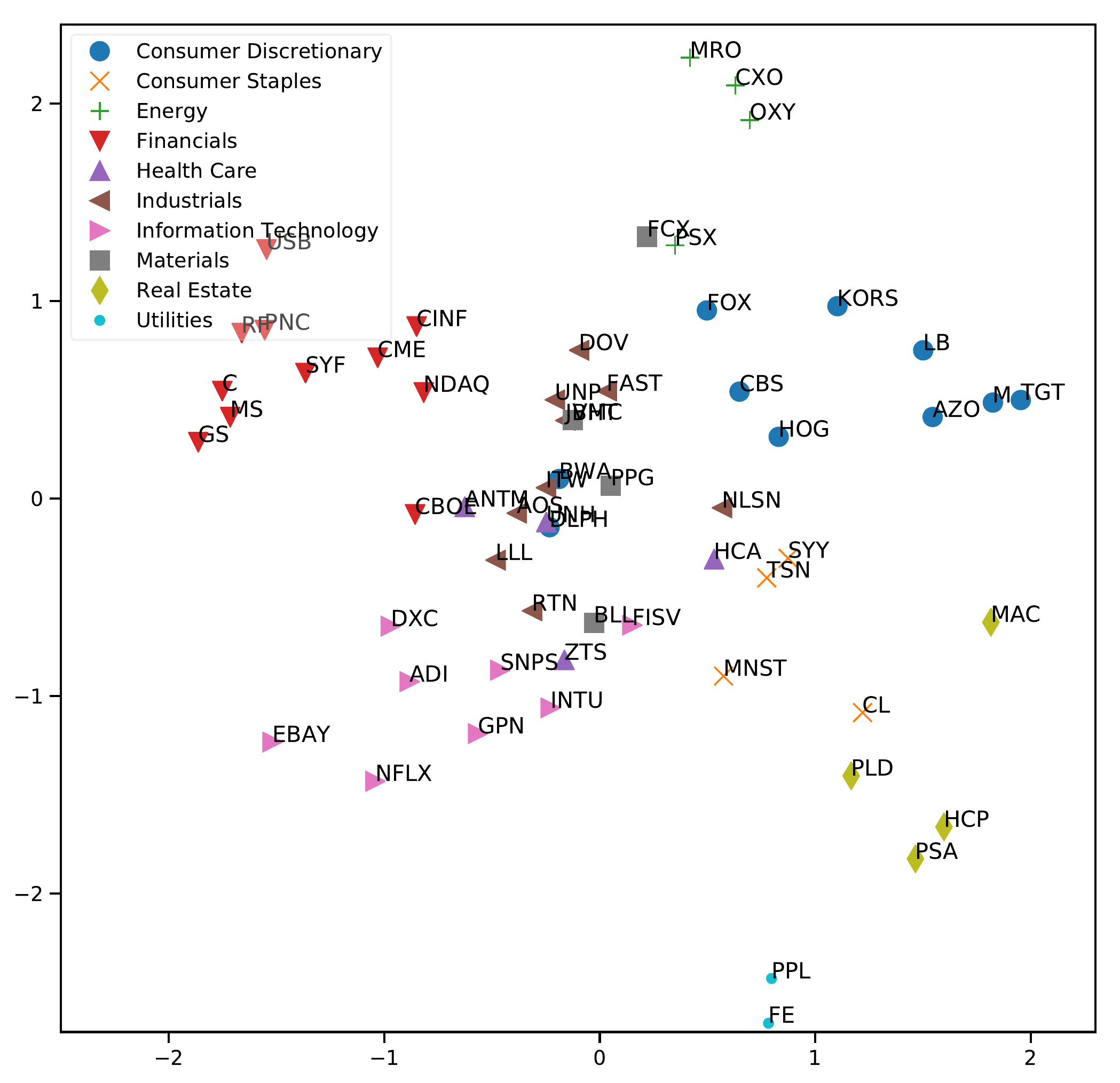}
 \caption{Stocks visualised in the 2-D latent space for the SE kernel.}
 \label{scatter_X}
\end{figure}

The 2-dimensional latent space can be visualized as a scatter plot.
For a stationary kernel function like the SE, the distance between the
stocks is directly related to their correlation. In this case, the latent positions
are even easier to interpret than market betas. As an example, 
Figure \ref{scatter_X} shows the 2-D latent space from 
60 randomly selected stocks from the S\&P500 from Jan 2017 to Dec 2017.
Visually stocks from the same sector tend to cluster together and we consider
our method as an alternative to other methods for detecting structure in financial
data \citep{Tumminello2010}.

\section{Conclusion}
\label{conclusion}

We applied the Gaussian Process Latent Variable Model (GP-LVM)
to estimate the covariance matrix between different assets, given their
time series. We then showed how the GP-LVM can be seen as a non-linear 
extension to the CAPM with latent factors. 
Based on the $R^2$-score and the ELBO, we concluded,
that for fixed latent space dimension $Q$, every non-linear kernel 
can capture more structure than the linear one.

The estimated covariance matrix helps us to build a minimal risk 
portfolio according to Markowitz Portfolio theory.
We evaluated the performance of different models on the S\&P500 from
year 2008 to 2018. Again, non-linear kernels had lower risk in the 
suggested portfolio and higher Sharpe ratios than the linear kernel and
the baseline measures. 
Furthermore, we showed how to use the GP-LVM to fill in missing prices
of less frequently traded assets and
we discussed the role of the latent positions of the assets.
In the future, one could also put a Gaussian Process on the latent positions
and allow them to vary in time, which would lead to a time-dependent 
covariance matrix.

\subsubsection*{Acknowledgments}

The authors thank Dr. h.c. Maucher for funding their positions.

\bibliographystyle{unsrt}
\bibliography{intelliSys2019}

\end{document}